\documentclass[10pt,a4paper,tightenlines]{article}
\usepackage{amsmath}
\usepackage{amsfonts}
\usepackage{amssymb}
\usepackage{amsthm}
\usepackage{placeins}
\usepackage{changepage}
\usepackage{graphicx}
\usepackage{xspace}
\usepackage{hyperref}
\usepackage{booktabs}
\usepackage{chicago}
\usepackage{url}
\usepackage{tcolorbox}

\newcommand{\gr}{``Green"\xspace}
\newcommand{\gw}{\gr-washing\xspace}
\newcommand{\xca}{XCA\xspace}

\theoremstyle{definition}
\newtheorem{manifesto}{Declaration}
\newtheorem{manifesto2}{Implication}

\begin{document}
\title{Sustainability Manifesto for Financial Products: Carbon Equivalence Principle\footnote{Chris Kenyon is Honorary Associate Professor at UCL.  The views expressed in this presentation are the personal views of the author and do not necessarily reflect the views or policies of current or previous employers. Not guaranteed fit for any purpose. Use at your own risk. Contact: c.kenyon@ucl.ac.uk. Mourad Berrahoui is Executive MBA student at Henley Business School.  The views expressed in this presentation are the personal views of the author and do not necessarily reflect the views or policies of current or previous employers. Not guaranteed fit for any purpose. Use at your own risk. Contact: ws802254@student.reading.ac.uk. Andrea Macrina is Reader in Mathematics at UCL. Contact: a.macrina@ucl.ac.uk.}}

\author{Chris Kenyon$^{\dag}$, Mourad Berrahoui$^{\ddag}$, Andrea Macrina$^{\dag\,\S}$
\\ \\ 
{$^{\dag}$Department of Mathematics, University College London} \\ {London, United Kingdom} 
\\ 
{$^{\ddag}$Henley Business School, University of Reading}\\{Reading, United Kingdom}
\\ 
{$^\S$African Institute of Financial Markets \& Risk Management} \\ {University of Cape Town, Rondebosch, South Africa} 
}
\date{7 December 2021\vskip10mm version 1.00}

\maketitle
\begin{abstract}
Sustainability is a key point for financial markets and the label \gr is an attempt to address this.  Acquisition of the label Green for financial products carries potential benefits, hence the controversy and attractiveness of the label.  However, such a binary label inadequately represents the carbon impact---we use carbon as a useful simplification of sustainability.  Carbon impact has a range either size of zero.   Both carbon emissions, and sequestration of carbon, are possible results of financial products.  A binary label does not allow differentiation between a carbon neutral investment and a coal power plant.  Carbon impact has timing and duration, a planted forest takes time to grow, a coal power plant takes time to emit.  Hence we propose the Carbon Equivalence Principle  (CEP) for financial products: that the carbon effect of a financial product shall be included as a linked term sheet compatible with existing bank systems.  This can either be a single flow, i.e., a summary carbon flow, or a linked termsheet describing the carbon impacts in volume and time.  The CEP means that the carbon impact of investment follows the money.  Making carbon impacts consistent with existing bank systems enables direct alignment of financial product use and sustainability, improving on non-compatible disclosure proposals.
\end{abstract}
	
\newpage
\tableofcontents

%=========================================================
\section{Introduction}
Sustainability is a key issue for financial market participants from investors to regulators.  One way to address sustainability has been the use of the label \gr with the potential to change prices and behavior \cite{tang2020shareholders}.  This label can be applied to any financial market product which makes the acquisition of this binary indicator attractive and controversial. Hence \gw is a systematic concern \cite{delmas2011drivers,alogoskoufis2021climate}.  However, we claim that the root issue is that a binary label is not fit for purpose for enabling sustainability: a binary label has no range---unlike carbon impacts, and a binary label has no time dimension---unlike carbon impacts.  We propose the Carbon Equivalence Principle (CEP): the carbon effects of a financial product shall be included as a linked termsheet to the product's existing termsheet.  The binary \gr label  is replaced by the impact in terms of equivalent carbon volume and time impacts.  Using a carbon termsheet compatible with existing bank systems, produces transparency enabling alignment of financial products and sustainability.  We complement disclosure proposals with related objectives \cite{fsb2017crfd,anderson2019ifrs,pra2019ss319}. Our proposal of compatibility with existing bank systems, means that all standard banking mechanisms can also be applied to carbon flows.

The shorthand version of the Carbon Equivalence Principle is that a financial product contains a single summary carbon flow of the carbon impact of the product.  The standard version of the CEP is that any financial product has two linked termsheets: one for the financial effects in terms of existing currencies or commodities, e.g., USD or XAU, and one for the carbon impacts. This adds, \xca\footnote{We use \xca for the carbon impact following the usual 3-letter guidelines: X is for non-national currencies, like XAU is for gold.  C is for carbon, and A is to get to three letters.} (carbon impact) to the list of currencies.    Usual position-keeping systems can handle trades split into several termsheets by linking them with a strategy identifier that downstream systems can then consume.

Consider the binary issue: What percentage saving of a single carbon molecule is required for a bond with notional of USD 100 million to qualify as \gr, and according to whom?  Alternatively, if the same bond saves one giga-ton of carbon emissions, does it have the same value?  The BIS analysis of sustainable finance taxonomies made the point \cite{ehlers2021taxonomy} as Principle 5, ``Sufficient Granularity'', and also proposed carbon emissions as the basis for a sustainable investment fund taxonomy.  We go further and propose the removal of the label \gr in financial products, replacing it by equivalent carbon impact: increase {\it and} decrease, and expand to cover the time dimension where the impacts actually occur.

In the retail world, the presence of sugar in food is not signaled by a binary label ``Sugared'' when there is sugar, but by the amount contained.  There is a label for the zero-sugar state ``Sugar-Free'', or ``Zero-Calorie'' and this makes sense because it is a significant state {\it and there is no possibility of negative sugar}.  In the case of carbon, there have been reoccurring proposals for carbon labelling at the retail level \cite{upham2011carbon}.

Now for the time issue of a binary label linked to the lifespan of a bond: Suppose that an investor buys a 15-year bond financing a coal power plant. A coal power plant has a lifespan of 35-50 years \cite{cui2019quantifying}; what about the emissions after the bond is repaid?  Consider a 15-year bond financing the planting of a forest.  A forest can absorb carbon for, say, 50-100 years before saturation \cite{pugh2019role}.  What about the benefits after the bond is repaid?  We emphasize that we propose only labelling, not monetization.  Monetization is a separate step, out of scope of this paper.  Of course \xca can be handled like any other currency (XAU, USD, etc) in booking systems. Equally, the termsheet in \xca is not qualitatively different from any other termsheet.  

\begin{figure}[th]
		\begin{center}
			\includegraphics[trim=0 0 0 0, clip, width=0.95\textwidth]{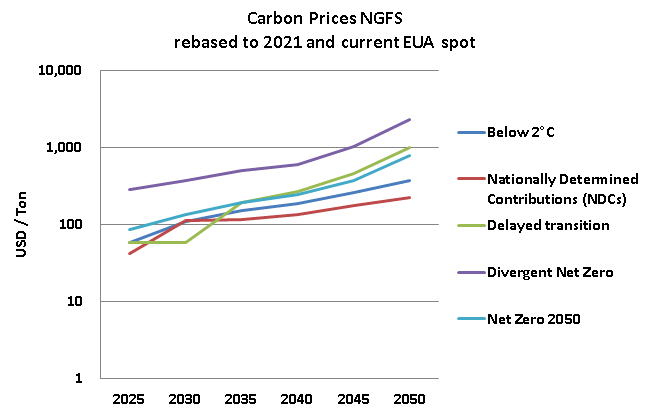}
		\end{center}
		\caption{Carbon price projections.  Source data (c) IIASA 2021, adapted for 2021 dollars from 2010 dollars, and current 2021 spot EUA price.  Used in accordance with licensing \url{https://data.ene.iiasa.ac.at/ngfs/\#/license}.}
		\label{f:ngfs}
\end{figure}

We are not proposing that \xca have any direct financial impact.  It is for the market to assign an exchange rate, or hedging strategy, if it so wishes.  This could be in terms of regulatory carbon emission permits (e.g., EUA \url{https://www.theice.com/products/197/EUA-Futures}), if it is in the scope of the regulation.  Alternatively, the voluntary market may provide unregulated offsets.  COP26 envisages expansion of voluntary markets under its non-official decisions with respect to Article 6 of the Paris Agreement \cite{agreement2015paris}.  However, carbon pricing is important and Figure \ref{f:ngfs} illustrates world-level carbon pricing according to NGFS scenarios adapted to 2021 dollars and EUA spot pricing.  We leave pricing for future work but emphasize that transparency is a required first step.

Here we propose a \emph{Sustainability Manifesto} for financial products based on the Carbon Equivalence Principle, and provide analysis and examples.  The objective of this manifesto is to enable and promote sustainability by providing transparency of the carbon impact produced by financial products in quantity and time, in a way that is compatible with existing bank systems.

Having \xca termsheets for financial products, in fact, is a bank requirement for those banks with net zero strategies.  Visibility is a requirement for them to be able to meet these targets transparently.  Clear measurement is a requirement for management,

The contributions of this paper are: firstly, the \emph{Carbon Equivalence Principle} enabling incentive alignment with sustainability in financial markets by providing transparency of carbon impact via termsheets compatibility with existing bank systems.  Secondly, we propose removing the ``Green" label to remove the representation and  \gw problems associated with a binary label.  Thirdly, we provide examples and operational details of handling the termsheets linked to \xca.

\section{Sustainability manifesto for financial products}

We first give the \emph{Sustainability Manifesto} and then the main implications.

\subsection{Carbon Equivalence Principle (CEP): \xca termsheets}
\begin{tcolorbox}
\begin{manifesto}{Carbon Equivalence for Financial Products I.}  The carbon impact of cashflows shall be linked to the causative cashflows.
\end{manifesto}
\begin{manifesto}{Carbon Equivalence for Financial Products II.}  All financial products shall include their equivalent carbon impact as part of their definition in the same way as the other parameters that define termsheets.  These termsheets shall be compatible with existing bank position-keeping systems.
\end{manifesto}
\begin{manifesto}{No ``Green'' label.}  The label ``Green'' may no longer be used for financial products.  
\end{manifesto}
\end{tcolorbox}

A bank position-keep system is the system used for storing financial product termsheets in electronic form that, at a minimum, enables life-cycle events to be tracked and acted upon.  Life-cycle events are, for example, payment or receipt of cashflows on specified dates.  These events can then give rise to changes in treasury cash positions.  Pricing and risk systems can be incorporated into position-keeping systems, or can be separate and just obtain positions (trades) from the position-keeping systems.

The labeling declaration can either be shorthand, i.e., a single flow, or standard---by which we mean that there is a linked termsheet for the carbon flows.

\subsection{Implications}
We now list the direct implications of the Sustainability Manifesto.
\begin{tcolorbox}
\begin{manifesto2}\emph{All financial products have linked \xca termsheets.}  The carbon impact of a product in quantity and time is treated in standard position-keeping systems, just like any other termsheet.  \xca termsheets are linked to the CCY (currency) termsheets using strategy identifiers enabling linked systems to process CCY and \xca impacts together, for pricing or risk management.
\end{manifesto2}
\begin{manifesto2}\emph{All financial products are linked to climate change.}  The carbon equivalence demonstrates that all financial products have environmental, i.e., climate impact, which increase or decrease atmospheric carbon. 
\end{manifesto2}
% later item on structuring captures changes to products
\begin{manifesto2}\emph{Product Maturity.}  The maturity of a financial product depends on both the non-\xca and \xca flows, and lifecycle events.
\end{manifesto2}
\begin{manifesto2}\emph{Transparency of Policy Choices.}  The price to make financial products net-zero reflects the policy choices of governments.  
\end{manifesto2}
\begin{manifesto2}\emph{Transparency of Policy Impact on Markets.}  The effect on markets from any change in price of carbon will be transparent.  
\end{manifesto2}
\begin{manifesto2}\emph{Price Discovery.}  Transparency of carbon impact enables price discovery of investor valuation.  A binary label is inadequate.  
\end{manifesto2}
\begin{manifesto2}\emph{Fungibility  and Structuring.}
Transparency of carbon-equivalent impact permits the carbon impact to be fungible, i.e., to be measured and priced uniformly for all products. Once the impact is fungible, then structuring is possible, i.e., addition and removal.
\end{manifesto2}
\end{tcolorbox}

Advantages and disadvantages of carbon equivalence principle for financial products are given in Table \ref{t:manifesto}.  Beyond solving the \gw problem by giving transparency to the climate impact, this enables {\bf fungibility} of the climate impact of the financial product.  

\begin{table}[h!]
\begin{adjustwidth}{-3cm}{-3cm}
\begin{center}
\begin{tabular}{p{4cm}p{2cm}p{3cm}p{5cm}l}\hline \\
\bf Feature &\bf ``Green'' label &\bf Carbon Equivalent &\bf Comments &\bf Advantage \\ \hline
\multicolumn{5}{c}{\bf Representation}\\
\hline
Shows absolute effect & no, relative &       yes   &  Shows actual effects & Carbon \\
Continuous range      & no, binary             &  yes                 &  Gradations of impact normal in reality      &     Carbon      \\
Include increases of carbon       &      no       &         yes          &  ``Green'' is relative  and cannot differentiate bad impacts     &          Carbon           \\
Includes decreases of carbon      &    yes: one        &   yes: range           & ``Green'' is relative          &           Carbon         \\ 
Can add impacts       &  no           &    yes               &    Investor can see overall {\bf net} impact      &     Carbon     \\ 
Simplicity        & too simple  & conceptually simple, operationally complex & binary is simplest possible but at cost of connection with reality & Carbon\\  
\hline
\multicolumn{5}{c}{\bf Pricing and Portfolio Implications}\\
\hline
Price discovery                    & inadequate & possible & a binary label is inadequate to enable price discovery of the value of the item & Carbon\\
\gw                                 &  susceptible  & difficult   &   Carbon equivalence makes climate impact  transparent               & Carbon  \\
Can price climate effect of portfolio       &       no      &       yes            &    Use carbon price over product life      &       Carbon   \\ 
Can separate product from climate effect       &  no           &    yes               &   Carbon equivalence allows separation and combination       &    Carbon    \\ 
Can achieve net zero climate effect of any portfolio by adding offsets       &   no          &   yes    &     Using ``Green'' label no way to calculate requirement       &  Carbon          \\ 
\hline
\end{tabular}
\label{t:manifesto}
\caption{Comparison of ``Green'' label with Carbon Equivalence.  The Advantage column compares the \gr label and Carbon Equivalence labelling to state which is better.}
\end{center}
\end{adjustwidth}
\end{table}

\subsection{CEP: shorthand examples}

Here we provide examples of how carbon impact could be attributed to different financial products.  This is not intended to be prescriptive or complete, but  a starting point. The carbon impact of a company can be classified into different scopes, typically direct, upstream and downstream.  Analysis of impact can start from direct and then broaden.
\begin{enumerate}
\item Equity and debt: Share of the carbon impact of company.
\item ESG-linked bond: Carbon impact of the specific project funded by the bond.  The bond may finance an ESG project and general expenditure.  In this case the carbon impact should be pro rata of the financing.
\item Bitcoin: Carbon impact to create and maintain over lifetime.
\end{enumerate}

\subsection{CEP: standard examples}

Here we give examples of project finance bonds with opposite effects: a forest-planting project; and a coal power plant example.  The numbers are functions of coefficient but the structure and qualitative aspects clearly show that the non-\xca currency flows can have a different maturity compared with the \xca flows.  This is the usual case for projects.  Some governments publish carbon emissions coefficients for power plants, e.g., Japan \cite{jp2021coef}.

\subsubsection{Linked termsheet: Forest-Planting Project}

Part 1: 10-year, fixed-rate USD bond financing project.
\begin{itemize}
    \item Notional is USD 100 million
    \item Cashflows are Modified Following 
    \item Party A pays Party B notional on 3 Jan. 2022
    \item Party B pays Party A notional on 3 Jan. 2032
    \item Party B pays Party A,  $x_1$\%\ annually on 29 Dec., first payment 2022, last payment 2032,  Act/360.
\end{itemize}
Part 2: \xca flows.
\begin{itemize}
    \item Notional is H hectares of planting area for tree type G
    \item Year 1: positive \xca H times $y_1$\%\ where $y_1$ is the planting cost in \xca
    \item Years 1 to 50: negative \xca, reverse amortization from zero in Year 1 to z\%\ times H in year 50.   
\end{itemize}

\subsubsection{Linked termsheet: Offshore Wind Power Plant}

We assume a one-year construction period and twenty years worth of operation, with subsequent one-year de-construction period (demolish and return to original state).

\vskip2mm
\noindent
Part 1: 21-year, fixed-rate USD bond that finances the wind power project.  
\begin{itemize}
    \item Notional is USD 100 million
    \item Cashflows are Modified Following.
    \item Party A pays Party B notional on 3 Jan. 2022
    \item Party B pays Party A notional on 3 Jan. 2043
    \item Party B pays Party A $x_2$\%\ annually on 29 Dec., first payment 2022, last payment 2043.  Daycount Act/360
\end{itemize}
Part 2: \xca flows.
\begin{itemize}
    \item Notional is M megawatts of electricity from offshore wind power plant, type K
    \item Year 1: positive \xca M times $y_2$\%\ where $y_2$ is the construction cost in \xca per megawatt
    \item Years 2 to 21: positive \xca M times $y_3$\%\ where $y_3$ is the maintenance cost in \xca per megawatt---expected to be very low
    \item Years 22: positive \xca M times $y_4$\%\ where $y_4$ is the de-construction cost in \xca per megawatt
\end{itemize}
A wind power plant may automatically be given carbon offsets by the relevant government.  In this case they will be part of the \xca budget.  In this case it is possible for the wind farm to be net zero.

\subsubsection{Linked termsheet: Coal Power Plant}

We assume a four-year construction period, and 35 years worth of operation, with subsequent four years de-construction period (demolish and return to original state).

\vskip2mm
\noindent
Part 1: 20-year, fixed-rate USD bond financing coal project.  
\begin{itemize}
    \item Notional is USD 1 billion
    \item Cashflows are Modified Following
    \item Party A pays Party B notional on 3 Jan 2022
    \item Party B pays Party A notional on 3 Jan 2042
    \item Party B pays Party A $x_3$\%\ annually on 29 Dec., first payment 2022, last payment 2042.  Daycount Act/360
\end{itemize}
Part 2: \xca flows.
\begin{itemize}
    \item Notional is M megawatts of electricity from coal power plant, type K
    \item Years 1 to 4: positive \xca M times $y_5$\%\ where $y_5$ is the construction cost in \xca per megawatt
    \item Years 4 to 35: positive \xca M times $y_6$\%\ where $y_6$ is the running cost in \xca per megawatt, expected to be high
    \item Years 35 to 39: positive \xca M times $y_7$\%\ where $y_7$ is the de-construction cost in \xca per megawatt
\end{itemize}

\section{Operational considerations}

Here we cover some of the operational considerations for financial products with termsheets linked \xca.

\subsection{Discounting \xca in the future and accumulating from the past}

If all \xca flows are offset exactly at the times when they occur, then the shorthand \xca value of the termsheet plus offsets will be zero.  If \xca flows are not offset, then to get from a \xca termsheet to a shorthand number one needs to move \xca flows in time, and in both directions.  So we need a physical discount rate for \xca.

Past \xca will accumulate and future \xca flows need to be brought back (i.e, discounted) to the present. Atmospheric carbon takes between to and twenty centuries to equilibrate via natural processes to 20\%-35\% of the original amount \cite{archer2009atmospheric}.  So, the discount rate for past cashflows is -2 to -35 basis points.  Future \xca flows have not started to be removed, so these use a discount factor of one. It may be that atmospheric carbon has non-linear effects on climate change after reaching certain barrier levels.  Here, we ignore these for simplicity, and because including them would add another modeling layer.  We leave this for future work.

It may be that atmospheric carbon has non-linear effects on climate change after reaching certain barrier levels.  We ignore these for simplicity, and because including them would add another modeling layer.  We leave this for future work.
\subsection{\xca currency-commodity-product complex and hedging}

\xca is the carbon currency, and there are a range of linked commodities and products.  For example, the Intercontinental Exchange trades EUA futures and options that have units of \xca.  The difference in electricity price at a location between producers with different levels of carbon emissions, prices a carbon spread between those emission levels.

\xca flows can be hedged by opposite \xca flows, or by regulatory or non-regulatory emissions allowances depending on the policy of the institution and the applicable regulatory regime.

\subsection{Linked termsheet lifecycle events}

We consider some simple examples of events during a bond lifecycle.

\subsubsection{Counterparty identification for \xca flows}

Using CEP, the \xca flows are from the party increasing or decreasing the carbon to the party providing the money for the change.  This makes accountability transparent. Emissions permits are options on future \xca flows between a buyer and the government, or government agency.  The permit can be used within a time-window, so have American-style optionality in time.

\subsubsection{Bond maturity}

In the case of a bond, when the bond is paid off, the future \xca flows move in the opposite direction to the money, i.e. back to the original issue who has now paid off the bond.

\subsubsection{Bond default}

If a bond defaults it is not paid off, so the \xca flows remain with the buyer of the bond.  However, if the bond defaults it is also likely that the \xca flows do not happen.  However, this is not guaranteed, it depends on the circumstances.

Equally, if for whatever reason the \xca flows are not honoured, then that is a default event just like non-payment of a CCY coupon.

\subsection{Extensions}

Here we mention some direct extensions of the CEP.  Note that we leave financial modeling of \xca prices for a separate, forthcoming paper.  The emphasis here is on the CEP manifesto and on showing how it can work.

\subsubsection{Floating \xca coupons}

Above, we have considered fixed \xca coupons.  In practice, many \xca coupons will be floating because the precise \xca impact will not be known until it is measured.  We can imagine a range of floating coupons depending on the \xca source, from forests to construction types, etc.  This additional information can also be captured in the linked termsheets.

\section{Alternative metrics}

We could go one step further and consider the effect of the carbon on the climate, and then propose pico-degrees Celsius as the labelling unit (pico = 10E-12).  According to \cite{arias2021climate}  1,150 giga tons of carbon are the limit for a two-degree change in temperature.  Thus, one ton of carbon is roughly one pico degree Celsius.  Moving to degrees requires an additional step from carbon, via models, to effects.  There are many models, and predictions vary significantly.  We pick the more conservative choice and stop at carbon equivalence.  For retail investors this step may be useful, we leave this for further investigation.

Alternatively we could propose a ratings system of discrete steps.  This could be designed to have negative impact ratings and positive impact ratings.  However, this is still inadequate at the portfolio level---how can an investor do arithmetic on the ratings of the bonds they hold to get a summary picture?  Furthermore, any discrete ratings system loses information because it is discrete.  Increasing the number of steps removes the apparent simplicity but retains the problem of getting an overall picture.

Carbon impact is only one dimension of sustainability, but arguably it is the most important.  For now, we pick a single dimension for simplicity.

\section{Conclusions}

The Carbon Equivalence Principle enables incentive alignment with sustainability in financial markets by providing transparency of carbon impact via compatibility with existing bank systems.  Removing the ``Green" label addresses the representation and  \gw problems associated with a binary label.  

We propose a shorthand CEP to label all financial products by equivalent carbon impact as a single flow: increases and decreases.  We also propose full transparency of impact via linked \xca termsheets in standard position-keeping systems to track carbon and cash over the lifetime of financial products.  

The  lifetime of a financial product can be considerably longer in \xca flows than in CCY cashflows.  This adds another consideration for valuation adjustments with respect to default, i.e., XVA on \xca.  %We leave XVA and pricing for further work.

\section{Acknowledgments}

The authors would like to gratefully acknowledge discussions with Susumu Higaki, Cathryn Kelly, Seiya Tateiri, and Kohei Ueda.

\bibliographystyle{chicago}
\bibliography{manifesto}

\end{document}